\documentclass[12pt, prd, showpacs]{revtex4}

\usepackage{amsmath}
\usepackage{amssymb}

\begin{document}

\title{The mass formula for quasi-black holes}
\author{Jos\'{e} P. S. Lemos}
\affiliation{Centro Multidisciplinar de Astrof\'{\i}sica, 
CENTRA, Departamento de F\'{\i}sica, 
Instituto Superior T\'ecnico - IST, Universidade T\'{e}cnica 
de Lisboa - UTL, Av. Rovisco Pais 1, 1049-001 Lisboa, Portugal\,\,}
\email{lemos@fisica.ist.utl.pt}
\author{Oleg B. Zaslavskii}
\affiliation{Astronomical Institute of Kharkov V.N. Karazin National
University, 35
Sumskaya St., Kharkov, 61022, Ukraine\,\,}
\email{ozaslav@kharkov.ua}

\begin{abstract}
A quasi-black hole, either non-extremal or extremal, can be broadly
defined as the limiting configuration of a body when its boundary
approaches the body's quasihorizon. We consider the mass contributions
and the mass formula for a static quasi-black hole. The analysis
involves careful scrutiny of the surface stresses when the limiting
configuration is reached. It is shown that there exists a strict
correspondence between the mass formulas for quasi-black holes and
pure black holes. This perfect parallelism exists in spite of the
difference in derivation and meaning of the formulas in both
cases. For extremal quasi-black holes the finite surface stresses give
zero contribution to the total mass. This leads to a very special
version of Abraham-Lorentz electron in general relativity in which the
total mass has pure electromagnetic origin in spite of the presence of
bare stresses.
\end{abstract}

\maketitle




\section{Introduction}

When the size of a compact body approaches its own gravitational
radius, usually the pressure, naturally assumed finite, cannot support
the body itself anymore, and gravitational collapse starts with some
margin before the gravitational radius is attained. Concurrently, it
turns out that there exist systems which possess static configurations
as close to the gravitational radius as one likes (see \cite{qbh} and
references therein).  They are called quasi-black holes and they
possess a would-be horizon, called a quasihorizon, instead of an event
horizon as for black holes. A quasi-black hole represents a particular
kind of a black hole mimicker, configurations close to be black holes
but having no event horizon \cite{mim}. Typical properties of
quasi-black holes consist in that the values of the lapse function on
the boundary and everywhere inside it tend to zero, giving rise to
whole regions of infinitely large redshifts. Such systems can be found
in quite different contexts, namely, self-gravitating monopoles,
extremal charged dust, either compact or dispersed
\cite{bon-d,extdust}, extremal charged shells \cite{shell}, and shells
gluing Reissner-Nordstr\"{o}m and Bertotti-Robinson spacetimes
\cite{glue1,glue2} (see also \cite{mim}).  Quasi-black holes, with
finite stresses, should be extremal, where for extremal one means that
the mass $M$ of such objects equals the charge $Q$, $M=Q$
\cite{qbh}. For a static object, this typically requires electric
charge, or some other form of appropriate repulsive charge.

Now, we want to extend the definition of quasi-black holes. In
\cite{qbh}, we indeed proved a theorem showing that quasi-black holes
should be extremal. This theorem was proved under the restrictive
assumption that the stresses on the surface of the body, the surface
stresses, on the quasihorizon should remain finite. Now, we want to
drop this assumption and allow for unbounded surface stresses on the
quasihorizon. Without restricting the surface stresses to be bounded
on the quasihorizon, there are certainly many more different types of
configurations that can achieve a quasihorizon. One can have now, also
non-extremal objects, as well as the extremal ones of the previous
considerations \cite{qbh}. So, a quasi-black hole, can be broadly
defined as the limiting configuration of a body, either non-extremal
or extremal, when its boundary approaches the body's quasihorizon.  We
will also include in our discussion ultraextremal quasi-black holes,
where the metric functions have a special behavior, related to the
extremal case but distinct \cite{rom,vo,tr}.  This ultraextremal
behavior also arises within black hole or cosmological solutions. For
instance, in the Reissner-Nordstr\"{o}m-de Sitter solution, an
ultraextremal triple horizon forms due to the existence of a special
relation between mass, charge and cosmological constant.

This enlargement in the definition of quasi-black holes will prove
crucial in the derivation of a quite general mass formula for the
quasi-black holes themselves, the main aim of this work. Indeed, our
analysis shows that there is a non-trivial connection between two
pairs of two different issues: (i) between surface stresses and the
mass formula for quasi-black holes, and (ii) between the mass formula
for quasi-black holes and the well known mass formula for black
holes. Concerning point (i) the ability to allow for, not only finite,
but also infinite stresses enlarges the spectrum of objects and gives
rise naturally to a mass formula. Concerning (ii) we show the close
correspondence between the mass formulas for quasi-black holes and
pure black holes both in the non-extremal and extremal cases. We want
to emphasize that both, the physical nature of the objects and the
derivation of the mass formula, is quite different for quasi-black
holes and black holes in turn. This certainly makes the close
relationship between the mass formulas non-trivial. Our analysis has
also rather unexpected consequences for the general relativistic
counterpart of the classical model of the Abraham-Lorentz
electron. These features are related with the distinguished role
played by the quasihorizon in the extremal case.

\section{Metric form and extension of the notion of static quasi-black
holes to encompass non-extremal cases}

\subsection{Metric form and definition of a quasi-black hole embodying
the non-extremal case}
\label{generalform}

Let us have a distribution of matter in a gravitational field which
does not depend on time. Put the four-dimensional spacetime metric
$ds^{2}=g_{\mu \nu}dx^{\mu }dx^{\nu }$, with $\mu,\nu$ being
spacetime indices, in the form
\begin{equation}
ds^{2}=-N^{2}dt^{2}+g_{ik}\left( dx^{i}+N^{i}dt\right) \left(
dx^{k}+N^{k}dt\right) \,,  \label{metricgeneral}
\end{equation}
where, we use $0$ as a time index, and $i,k=1,2,3$ as spatial indices. In
addition, $N$ and $N^{i}$ are the lapse function and shift vector which
depend in general on the spatial coordinates $x^{i}$. Putting $N^{i}=0$ to
study the static case, the metric (\ref{metricgeneral}) reduces to
\begin{equation}
ds^{2}=-N^{2}dt^{2}+g_{ik}dx^{i}dx^{k}\,,  \label{metricstatic0}
\end{equation}
where $N$ is a function of the spatial coordinates. It is further
convenient to work in Gauss normal coordinates where the metric looks
like (see \cite{vis}, and e.g., \cite{adler})
\begin{equation}
ds^{2}=-N^{2}dt^{2}+dl^{2}+g_{ab}dx^{a}dx^{b}\,,  \label{metricstatic}
\end{equation}
with $l$ being a radial coordinate, and $a,b$ representing the other
two spatial coordinates. For instance, if the metric is spherically
symmetric they are the angular coordinates $\theta$ and $\phi$.  Note
that Gaussian normal coordinates cannot be extended beyond the point
where geodesics normal to the surface begin to form caustics. However,
we are interested in the vicinity of the body's surface only, which is
going to become a quasihorizon, so for our purposes it is quite
sufficient to use the reference system (\ref{metricstatic}) in this
vicinity.

We suppose that the body is compact with its boundary approaching the
quasihorizon \cite{bon-d} or, as in \cite{extdust}, the distribution
can be dispersed but with a well-defined quasi-black hole limit. We
consider the static case, and assume no further symmetry, spherical or
whatever.  Previously, the definition of a quasi-black hole, and the
corresponding quasihorizon, was done in \cite{qbh} for spherically
symmetric spacetimes.  Its generalization to static spacetimes without
the requirement of spherical symmetry leads to the following
points. Consider a configuration depending on a parameter
$\varepsilon$ such that (a) for small but non-zero values of
$\varepsilon$ the metric is regular everywhere with a non-vanishing
lapse function $N$, at most the metric contains only delta-like
shells, (b) taking as $\varepsilon$ the maximum value of the lapse
function on the boundary $N_{\mathrm{B}}$, then in the limit
$\varepsilon \rightarrow 0$ one has that the lapse function $N\leq $
$N_{\mathrm{B}}\rightarrow 0$ everywhere in the inner region, (c) the
Kretschmann scalar $\mathrm{Kr}$ remains finite in the quasihorizon
limit.  This latter property implies another important property which
can be stated specifically, namely, (d) the area $A$ of the
two-dimensional boundary $l=\mathrm{const}$ attains a minimum in the
limit under consideration, i.e., $\lim_{\varepsilon \rightarrow
0}\frac{\partial A}{\partial l}|_{l^{\ast }}=0$, where $l^{\ast }$ is
the value of $l$ at the quasi-horizon.  When a configuration obeys
these three properties (a)-(c) (or (a)-(d)) we say one is in the
presence of a quasi-black hole, enlarging to non-spherically symmetric
spacetimes, though still static, the definition given in \cite{qbh}.
A remark should be made: In many cases of physical interest,
especially for the spherically symmetric systems, the lapse function
is a monotonically decreasing function of the proper distance in the
direction from the boundary to the inner region (see Appendix B in
\cite{qbh}). Then, we can weaken point (b) and require only that the
maximum boundary value obeys $N_{\mathrm{B}}\rightarrow0$.

\subsection{Finiteness of the Kretschmann scalar: Elaboration of
property (c)}
\label{finiteness}

Let us elaborate on property (c). In \cite{qbh} only extremal
configurations were considered. As discussed in \cite{qbh}, these are
regular in the sense that the Kretschmann scalar $\mathrm{Kr}$ remains
finite in the quasihorizon limit (as property (c) above demands), and
in addition the surface stresses (if any) also remain regular in that
limit. Now, for the non-extremal case the attempt to use the same
notion for a quasi-black hole leads to infinite stresses as shown in
\cite{qbh}, so that one should allow for infinite stresses if one
wants to include the non-extremal case. Therefore one should also ask
whether or not for non-extremal quasi-black holes one should insist,
as in (c) above, that $\mathrm{Kr}$ remains finite. We showed in a
previous article \cite{mim} that, typically, point (c) can be violated
for mimickers, generic configurations close to be black holes but
having no event horizon, of which a quasi-black hole is an
example. For such singular configurations the notion of a quasi-black
hole, non-extremal one, would be devoid of meaning. Therefore, to keep
non-extremal quasi-black holes as physically relevant objects, we
demand that point (c) should be maintained in the list of properties
above as an important requirement.

\subsubsection{Non-extremal case}

We will now see what consequences property (c) by looking at the
explicit expression for $\mathrm{Kr}$. We will follow \cite{vis}
closely, where true black holes, rather than quasi-black holes, were
considered. One can obtain from (\ref{metricstatic}) that the
Kretschmann scalar $\mathrm{Kr}$ is given by
\begin{equation}
\mathrm{Kr}=P_{ijkl}P^{ijkl}+4C_{ij}C^{ij}\,,  \label{Kr}
\end{equation}
where $P_{ijkl}$ is the curvature tensor for the subspace
$t=\mathrm{const}$, and
\begin{equation}
C_{ij}=\frac{N_{\mid ij}}{N}\,,  \label{Kr2}
\end{equation}
with $_{\mid i}$ denoting the covariant derivative with respect to the
corresponding three-dimensional metric. As the metric of the
three-space is positive definite, all terms enter the entire
expression (\ref{Kr}) with a positive sign, so that each term should
be finite separately.

For instance, let us see the pure black hole case as done in
\cite{vis}. The finiteness of $\mathrm{Kr}$ entails that in the
horizon limit, when $N\rightarrow 0$, the numerator in $C_{ij}$ (see
Eq. (\ref{Kr2})) must vanish. Without loss of generality, for the
non-extremal horizon one can choose $l=0$. Considering then different
combinations of indices, one arrives at the asymptotic form of $N$ for
the pure black hole case \cite{vis},
\begin{equation}
N=\kappa l+\kappa _{3}\frac{l^{3}}{3!}+O(l^{4})  
\label{bh}
\end{equation}
where $\kappa=\mathrm{const}$ is the surface gravity of the black
hole, and $\kappa_{3}$ is some function.

Now, we want to study the quasi-black hole case instead of the black
hole case. Choosing the coordinate $l$ in such a way that $l=0$ on the
boundary surface, where $N=N_{0}(x^{a})\rightarrow 0$, in the limit
under discussion we obtain that
\begin{equation}
\lim_{l\rightarrow 0}C_{ll}=\frac{\lim_{l\rightarrow 0}N^{\prime
\prime }}{ N_{0}}  
\label{11}
\end{equation}
where $\,^{\prime }\equiv \frac{\partial }{\partial l}$, and,
\begin{equation}
\lim_{l\rightarrow 0}C_{al}=\frac{\lim_{l\rightarrow 0}N_{;a}^{\prime
}}{ N_{0}}\,, 
\label{1a}
\end{equation}
where $\,_{;a}$ means the covariant derivative with respect to the
metric $g_{ab}$ in Eq. (\ref{metricstatic}). We can write
$N_{\mathrm{0} }=\varepsilon f(x^{a})$, with $\varepsilon =0$
corresponding to the quasi-black hole limit. Then, it follows from the
finiteness of $\mathrm{Kr}$ that
\begin{equation}
\lim_{\varepsilon \rightarrow 0}\lim_{l\rightarrow 0}
N^{\prime \prime }=0\,,
\label{Clim1}
\end{equation}
\begin{equation}
\lim_{\varepsilon \rightarrow 0}\lim_{l\rightarrow 0}N_{;a}^{\prime }=0\,.
\label{Clim2}
\end{equation}
If we write the expansion for small $l$ in the form
\begin{equation}
N=N_{\mathrm{0}}+\kappa _{1}(x^{a},\varepsilon )l+\kappa
_{2}(x^{a},\varepsilon )\frac{l^{2}}{2!}++\kappa
_{3}(x^{a},\varepsilon)
\frac{l^{3}}{3!}+O(l^{4})\,,  
\label{expansion}
\end{equation}
and take into account (\ref{Clim1})-(\ref{Clim2}) we obtain that
\begin{equation}
\lim_{\varepsilon \rightarrow 0}\kappa _{1}(x^{a},\varepsilon )=\kappa
\label{surfgrav}
\end{equation}
is a constant, and
\begin{equation}
\lim_{\varepsilon \rightarrow 0}\kappa _{2}=0.  \label{k2}
\end{equation}
Thus, we see that the expansion (\ref{expansion}) has the same
structure as (\ref{bh}). From the meaning of a quasi-black hole, we
want in the outer region to have, $\lim_{\varepsilon \rightarrow
0}N(\varepsilon ;l,x^{a})=N_{ \mathrm{bh}}(l,x^{a})$, where
$N_{\mathrm{bh}}$ is the lapse function for a black hole (this can be
not the case for the inner region, see \cite{qbh}).  Now, in
(\ref{surfgrav}) we wrote the limit is $\kappa $, but strictly
speaking we should have put $\kappa _{\mathrm{h}}$, the surface
gravity of the quasi-black hole. Therefore, the final step consists in
identifying indeed $\kappa_{\mathrm{h}}$ with the surface gravity
$\kappa$, so that we a have a well-defined limit for the quantity
$\kappa _{1}.$ Thus, as far as the properties of the metric are
concerned, we have proved that
\begin{equation}
\lim_{l\rightarrow 0}\lim_{\varepsilon \rightarrow 0}=\lim_{\varepsilon
\rightarrow 0}\lim_{l\rightarrow 0}.  \label{limits}
\end{equation}
For a configuration which is close to the quasi-black hole limit but
does not attain it, there are slight deviations of the coefficients
$\kappa_{1}$ and $\kappa_{2}$ from their limiting values but, the
closer to the limit is the configuration, the smaller the corrections
become. In the quasi-black hole limit one can ignore those corrections
altogether and consider, in particular, the surface gravity as a
constant, similarly to what happens to black holes. So, $\kappa
_{\mathrm{h}}=\kappa$. We would like to stress that the validity of
the expansion (\ref{expansion}) with the additional properties
(\ref{surfgrav})-(\ref{k2}) is an essential property. For example, in
\cite{mim} we considered black hole mimickers for which $N=\sqrt{
V+\varepsilon ^{2}}$, so that $\lim_{\varepsilon \rightarrow
0}\lim_{l\rightarrow 0}\frac{\partial ^{2}N}{\partial l^{2}}$ diverges
and the expansion (\ref{expansion}) fails. Such configurations have
singular limits, indeed $\mathrm{Kr}$ diverges. Thus, not any
deformation of a black hole metric depending on some deformation
parameter $\varepsilon$ (see \cite{mim}) is suitable if we want to
give a reasonable extension of the notion of quasi-black black holes
to the non-extremal case. The requirement (c) formulated above selects
then admissible deformations among the possible ones.

We show now how property (d) follows from the above requirements.
Similarly to the expansion for the lapse function $N$ (see
Eq. (\ref{expansion})), we can write the expansion for the metric
$g_{ab}$ as, \begin{equation}
g_{ab}=g_{ab}^{(0)}(x^{a})+g_{ab}^{(1)}(x^{a})l+\frac{g_{ab}^{(2)}}{2}
(x^{a})l^{2}+O(l^{3})\,.  \end{equation} Then, from the requirement of
the finiteness of $C_{ab}$ we obtain for the extrinsic curvature,
$K_{ab}$, define here as $K_{ab}=-\frac{1}{2}\frac{\partial
g_{ab}}{\partial n}$, that
\begin{equation}
\lim_{\varepsilon \rightarrow 0}\lim_{l\rightarrow 0}K_{ab}=0\,,
\end{equation}
similarly to the property $\lim_{l\rightarrow 0}\lim_{\varepsilon
\rightarrow 0}K_{ab}=0$ which is known to hold for black holes
\cite{vis}. Finally, for quasi-black holes, we obtain that the area of
the cross-section $l=\rm const$ obeys,
\begin{equation}
\lim_{\varepsilon \rightarrow 0}\lim_{l\rightarrow 0}
\frac{1}{A}\frac{
\partial A}{\partial l}=-\frac{1}{2}\lim_{\varepsilon \rightarrow
0}\lim_{l\rightarrow 0}K_{ab}=0\,,
\label{area}
\end{equation}
which is just property (d) mentioned at the end of Sec. 
\ref{generalform}.

\subsubsection{Extremal and ultraextremal cases}
\label{extrultra}

In the extremal case the situation is even simpler. Consider a
quasi-black hole in which a small parameter, $\varepsilon \neq 0$ say,
enumerates configurations. Then, by the definition of a quasi-black
hole, $\lim_{\varepsilon \rightarrow 0}N(l^{\ast
},x^{a};\varepsilon)=0$, where $l^{\ast }=l^{\ast}(\varepsilon)$
corresponds to the proper distance between any fixed point and the
quasihorizon \cite{qbh}. Here one comment is in order. Actually, in
relation to the non-extremal case, we use a somewhat different
definition of proper distance here. This simply reflects the
qualitatively different properties of the non-extremal and extremal
black hole geometries to which a corresponding quasi-black hole tends
in the outer region, in the limit $\varepsilon \rightarrow 0$.
Indeed, for a non-extremal black hole the proper distance from the
horizon to any other point is finite, so that without loss of
generality we have adapted $l$ so that $l=l^{\ast}=0$ for the
quasihorizon itself. In the extremal case the proper distance from the
horizon to any other point is infinite, so one has to measure $l$ not
from the horizon but from any other fixed point, $l\rightarrow
\infty$, when the second point approaches the
horizon. Correspondingly, for a quasi-black hole in the extremal case
the proper distance from a fixed point to a quasi-horizon $l^{\ast
}(\varepsilon )$ is finite but $\lim_{\varepsilon \rightarrow
0}l^{\ast }(\varepsilon )=\infty$, justifying thus our choice for the
extremal case. Now, we take into account that if a continuous
function, $f(x)$ say, of an arbitrary variable $x$, is such that the
limit $f_{\infty}=\lim_{x\rightarrow \infty }f(x)$ exists and is the
finite (roughly speaking, the function approaches asymptotically a
constant value), it follows that $\lim_{x\rightarrow \infty
}\frac{df}{dx}=0$. So, this means that $\lim_{\varepsilon \rightarrow
0}\frac{\partial N}{\partial l^{\ast }}=0$. It can be rewritten as
$\lim_{\varepsilon \rightarrow0}\left( \frac{ \partial N}{\partial
l}\right) _{\mathrm{h}}=0$, where the subscript h means here that the
quantity is calculated on the quasihorizon. If we take the limits in
the other order, we simply return to the usual black hole, which, by
definition, in the limit $l\rightarrow \infty$ the lapse function
behaves as $N\sim \exp (-Bl)$ with $B=\mathrm{const}>0$ in the
extremal case. For an ultraextremal quasi-black hole, one has that the
asymptotic behavior of the lapse function is given by $N\sim l^{-n}$,
with $n>0$, and the choice of $l^{\ast }$ is the same as for the
extremal case.  There are also black holes that have this
ultraextremal behavior, for example, the Reissner-Nordstr\"{o}m-de
Sitter solution with a triple horizon occuring due to a special
relation between mass, charge and cosmological constant
\cite{rom,vo,tr}.  The ultraextremal quasi-black hole and black hole
cases have the same correspondence between themselves as the extremal
cases.  Thus, in brief, the property that extremal and ultraextremal
black holes have zero surface gravity at their horizon has a
corresponding identical property for extremal and ultraextremal
quasi-black holes at their quasihorizon.

We show now, for the extremal and ultraextremal cases, how property
(d) follows from the above requirements.  The metric $g_{ab}$ in the
extremal case, say, has the following expansion,
\begin{equation}
g_{ab}=g_{ab}^{(0)}+g_{ab}^{(1)}\exp
(-\frac{l}{l_{0}})+g_{ab}^{(1)}\exp (- \frac{2l}{l_{0}})+...
\label{ge}
\end{equation}
where $l_{0}$ is a constant. The maximum value of $\,l$ is equal to
$l^{\ast}$, which corresponds to the value of the proper distance
between a fixed point and the quasi-horizon. For a finite parameter
$\varepsilon$ the quantity $l^{\ast}$ is also finite, but
$l^{\ast}\rightarrow \infty$ when $\varepsilon \rightarrow 0$. Then,
again, we obtain that the tensor $K_{ab}$ on the quasi-horizon, in the
limit $\varepsilon \rightarrow 0$, obeys $K_{ab}\rightarrow 0$.
Correspondingly, property (d) holds (note here that in
Eq. (\ref{area}) in the limiting process, the value $l=0$
corresponding to the quasi-horizon should be replaced by
$l=\infty$). In the ultraextremal case we have, instead of (\ref{ge}),
an expansion with respect to inverse powers of $l$ that also leads to
Eq. (\ref{area}), i.e., to property (d).

\subsubsection{The three cases together: Non-extremal, extremal and
ultraextremal}

Now, for extremal and ultra extremal cases, the surface gravity obeys 
$\kappa=0$ by definition. So, we can combine all three cases, i.e.,
non-extremal, extremal and ultraextremal, in the formula,
\begin{equation}
\lim_{\varepsilon \rightarrow 0}\left( \frac{\partial N}{\partial l}
\right)_{\mathrm{h}}=\kappa \,,  \label{k}
\end{equation}
where again the subscript h means here that the quantity is calculated
on the quasihorizon, and where $\kappa$ is equal to the surface
gravity of the corresponding black hole.  Roughly speaking, for the
outer region, a quasi-black hole represents an object that realizes
the limiting transition from an \textquotedblleft would-be black
hole\textquotedblright\ to a true one. Therefore, it is not surprising
that there is a direct correspondence between their features.

Let us conclude this section with some general remarks. Actually, the
properties (a)-(c) listed in Sec. \ref{generalform} mean that in the
limit $\varepsilon \rightarrow 0$ the metric of a quasi-black hole
approaches that of a black hole everywhere in the outer region (let us
stress again that this is not necessarily the case for the inner
region because of the complex entanglement between coordinates and
parameters in the course of the limiting process
\cite{qbh}). Therefore, it is quite trivial that far from the
quasi-horizon the derivatives of the metric also coincide in the limit
under discussion.  However, in the vicinity of the quasi-horizon,
because of the interplay between two small parameters, $\varepsilon $
and $l$ in the non-extremal case, or $\varepsilon $ and $1/l$ in the
extremal or ultraextremal cases, this is not obvious in advance, and
additional substantiation is necessary to establish Eq. (\ref{k}), as
was done in the above consideration. In addition, in all three cases
the property (d) holds.

\section{The mass formula for the generic static case}

\label{genericstaticcase}

If the matter is joined onto a vacuum spacetime then one has to be
careful and use the junction condition formalism \cite{isr,mtw}. The
mass of the matter distribution can be written as an integral over the
region occupied by matter and fields. Defining $T_{\mu }^{\nu }$ as
the stress-energy tensor, the mass is given by the Tolman formula
(see, e.g., \cite{p}, or for the original work \cite{tolman}, see also
\cite{ll}),
\begin{equation}
M=\int (-T_{0}^{0}+T_{k}^{k})\sqrt{-g}\,d^{3}x\,,  \label{mtot}
\end{equation}
where $g$ is the determinant of the metric $g_{\mu \nu }$. This is the
starting point of our analysis. We discuss this integral to find the
mass formula of a quasi-black hole. For the mass formula for black
holes, rather than quasi-black holes, see \cite{bch,cart73,bard73,sm},
and, e.g., \cite{fn}.

\subsection{The various masses}

\subsubsection{Total mass}

Since the spacetime is static by assumption, the distribution of
matter does not depend on time. Then using (\ref{mtot}) and noting
from (\ref{metricstatic}) that $\sqrt{-g}=N\sqrt{g_{3}}$, where
$g_{3}$ is the determinant of the spatial part of the metric
(\ref{metricstatic}), i.e., is the determinant of the metric on the
hypersurface $t=\mathrm{constant}$, one finds
\begin{equation}
M=\int (-T_{0}^{0}+T_{k}^{k})\,N\sqrt{g_{3}}\,d^{3}x\,.  
\label{mtotstatic1}
\end{equation}
The mass (\ref{mtotstatic1}) can be split into three different
contributions, from the inner region, the outer one (where, for example, a
long-range electromagnetic field, or other matter fields, such as rings, can
be present) and from the surface between the two,
\begin{equation}
M_{\mathrm{tot}}=M_{\mathrm{in}}+M_{\mathrm{surf}}+M_{\mathrm{out}}\,,
\end{equation}
where, $M_{\mathrm{in}}$, $M_{\mathrm{surf}}$, $M_{\mathrm{out}}$, are
the inner mass, the surface mass, and the outer mass,
respectively. Let us study each mass term in turn.

\subsubsection{Inner mass}
\label{innermasssection}

{}From Eq. (\ref{mtotstatic1}), the inner mass is given 
by the expression,
\begin{equation}
M_{\mathrm{in}}=\int_{\mathrm{inner}}(-T_{0}^{0}+T_{k}^{k})\,N\sqrt{g_3}
d^{3}x\,\,.
\label{inner}
\end{equation}
As we presume a quasi-black hole to form, it means that in the entire
inner region $N\leq N_{\mathrm{B}}$ where $N_{\mathrm{B}}$ is the
maximum boundary value and, as $N_{\mathrm{B}}\rightarrow 0$, also
$N\rightarrow 0$ everywhere in the inner region \cite{qbh}. Therefore,
one can write the following inequality for the inner mass,
$M_{\mathrm{in}}=\int_{\mathrm{\ inner }}(-T_{0}^{0}+T_{k}^{k})
\,N\sqrt{g_{3}}\,d^{3}x\leq N_{\mathrm{B}} \int
(-T_{0}^{0}+T_{k}^{k})\sqrt{g_{3}}\,d^{3}x$. Defining the proper mass
$ M_{0}$ as $M_{0}\equiv -\int \, T_{0}^{0}\sqrt{g_{3}}\,d^{3}x$, and
the mass due to the stresses as $M_{k}\equiv \int
\,T_{k}^{k}\sqrt{g_{3}}\, d^{3}x$, one finds
\begin{equation}
M_{\mathrm{in}}\leq N_{\mathrm{B}}\,(M_{0}+M_{k})\,.
\end{equation}
By assumption, the proper mass $M_{0}$ is finite. Assuming also that 
$T_{k}^{k}\leq C\left\vert T_{0}^{0}\right\vert$ where $C$ is some
constant, we obtain that $M_{k}$ is finite as well. Thus, in the
quasi-black hole limit,
\begin{equation}
M_{\mathrm{in}}= 0\,,  \label{minst}
\end{equation}
due to the factor $N_{\mathrm{B}}$.

\subsubsection{Surface mass}

Now consider the contribution from the surface,
\begin{equation}
M_{\mathrm{surf}}=\int_{\mathrm{surface}}(-T_{0}^{0}+T_{k}^{k})\,N\sqrt{
g_{3} }\,d^{3}x\,\,.
\end{equation}
Here, there are delta-like contributions in $T_{\mu }^{\nu }$, the
surface stresses being then given by,
\begin{equation}
S_{\mu }^{\nu }=\int T_{\mu }^{\nu }\,dl\,,
\end{equation}
where the integral is taken across the shell. Define $\alpha $ as,
\begin{equation}
\alpha =8\pi (S_{a}^{a}-S_{0}^{0})\,,  \label{a}
\end{equation}
so that from a combination of some of the equations above, we get,
\begin{equation}
M_{\mathrm{surf}}=\frac{1}{8\pi }\int \alpha \,N\,d\sigma \,,  
\label{ma}
\end{equation}
where $d\sigma =\sqrt{g_{2}}\,d^{2}x$, $g_{2}$ being the determinant
of the metric spanned by the $x^{a}$ (see
Eq. (\ref{metricstatic})). Now, one also has the relationship $8\pi
S_{\mu }^{\nu }=[[K_{\mu }^{\nu }]]-\delta _{\mu }^{\nu
}[[K]]\label{s0}$, where $K_{\mu }^{\nu }$ is the extrinsic curvature
tensor, $[[...]]=[(...)_{+}-(...)_{-}]$, and subscripts
\textquotedblleft $+$ \textquotedblright\ and \textquotedblleft
$-$\textquotedblright\ refer to the outer and inner sides,
respectively (see, e.g., \cite{isr,mtw}). Thus $ \alpha =8\pi
(S_{a}^{a}-S_{0}^{0})=-2[[K_{0}^{0}]]$. Put $n^{\mu}$ as the unit
vector normal to the boundary surface. Then $K_{0}^{0}=-{\ n^{0}}
_{;\,0}=- \frac{1}{N}\frac{\partial N}{\partial l}$. As a result, we
obtain
\begin{equation}
\alpha =\frac{2}{N}\left[ 
\left( \frac{\partial N}{\partial l}\right)
_{+}-\left( \frac{\partial N}{\partial l}\right) _{-}\right] \,,  
\label{an}
\end{equation}
and so,
\begin{equation}
M_{\mathrm{surf}}=\frac{1}{4\pi }\int_{\mathrm{surf}}\left[ 
\left( \frac{
\partial N}{\partial l}\right) _{+}-
\left( \frac{\partial N}{\partial l}
\right) _{-}\right] d\sigma \,.  
\label{al}
\end{equation}
This shows clearly that one cannot ignore the surface stresses in the
non-extremal case. This is a very important feature of non-extremal
configurations which can be confronted with the extremal ones. One
could na\"{\i}vely think that one could simply restrict oneself to the
case of vanishing stresses but in the problem under discussion this is
impossible.  Indeed, it is seen that the stresses enter the mass
formulas via the quantity $\alpha $, so in the case of vanishing
stresses $M_{\mathrm{surf}}$ would also vanish. But this is obviously
impossible in the non-extremal case. Indeed, in the quasi-black hole
limit, the situation we want to analyze in detail, one has
$(\frac{\partial N}{\partial l})_{-}\rightarrow 0$, since everywhere
in the inner region $N$ is bounded and tends to zero by the definition
of a quasi-black hole \cite{qbh}, so that $(\frac{\partial N}{
\partial l})_{-}\rightarrow 0$. Thus, in the limit, $\alpha
\rightarrow \frac{2}{N}(\frac{\partial N}{\partial l})_{+}$. Now,
according to (\ref{expansion}) and (\ref{surfgrav}), one has $\alpha
\simeq 2\frac{\kappa }{N}$, where $\kappa $ is the surface gravity. It
diverges since in general at the quasihorizon $\kappa \neq 0$ and
$N\rightarrow 0$. But the surface contribution to the mass is finite
due to the factor $N$. So,
\begin{equation}
M_{\mathrm{surf}}=\frac{\kappa A_{\mathrm{h}}}{4\pi }\,,  \label{ms}
\end{equation}
where $A_{\mathrm{h}}$ is the quasihorizon area. Eq. (\ref{ms}) is
valid in general, i.e., it is valid in the non-extremal case where
$\kappa \neq 0$, and in the extremal case where, since $\kappa =0$,
the surface mass is zero, $M_{\mathrm{surf}}=0$.

Let us study the extremal case in more detail. In the extremal case
one has, near the quasihorizon, $l\rightarrow\infty$ and $N\sim
\exp(-Bl)$ where $B$ is a constant. Therefore, $\alpha =-2B$ is finite
(for instance, for a system forming a quasi-black hole whose exterior
metric is Reissner-Nordstr\"{o}m, one has
$B=\frac{1}{r_{\mathrm{+}}}$, where $r_{\mathrm{+}}$ is the horizon
radius, and so, $\alpha =-\frac{2}{r_{\mathrm{+}}}$, finite). Then, it
follows from (\ref{al}) that $M_{\mathrm{surf}}=0$. The surface
stresses themselves are not equal to zero but they do not contribute
to the mass in the extremal case, since it is multiplied by the factor
$N$ in (\ref{ma}).  For completeness, we also mention the
ultraextremal case, with the asymptotic behavior of the lapse function
being $N\sim l^{-s}$, $s>0 $.  Then, it is seen from (\ref{al}) that
not only the contribution of the surface to the total mass vanishes,
but the surface stresses themselves vanish as well.

\subsubsection{Outer mass}

The outer mass is given by the expression,
\begin{equation}
M_{\mathrm{out}}=
\int_{\mathrm{out}}(-T_{0}^{0}+T_{k}^{k})\,N\sqrt{g_{3}}
d^{3}x\,.
\end{equation}
Further, we may split $M_{\mathrm{out}}$ into an electromagnetic part
$M_{\, \mathrm{out}}^{\mathrm{em}}$, and a non-electromagnetic part,
$M_{\;\mathrm{\ \ out}}^{\mathrm{matter}}$ say, for the case of dirty
black holes or dirty quasi-black holes, exactly in the manner as it
was already done in \cite{cart73}, and obtain
$M_{\mathrm{out}}=M_{\,\mathrm{out}}^{\mathrm{em} }+M_{\;
\mathrm{out}}^{\mathrm{matter}}$. Since $M_{\,\mathrm{out}}^{\mathrm{
\ em} }=\varphi _{\mathrm{h}}Q$, as explained below, where $\varphi _{
\mathrm{h }}$ is the electric potential on the horizon in the case of
black holes, and the electric potential on the quasihorizon in the
case of quasi-black holes, and $Q$ is the corresponding electric
charge, one finds
\begin{equation}
M_{\mathrm{out}}=\varphi _{\mathrm{h}}Q+
M_{\;\mathrm{out}}^{\mathrm{matter}}\,.  
\label{outerm}
\end{equation}

Now, we justify that $M_{\,\mathrm{out}}^{\mathrm{em}}=
\varphi_{\mathrm{h} }Q$. As is explained above, the inner contribution
of the matter to the mass vanishes (independently of the kind of
matter or field), provided the components of the stress-energy tensor
within the matter remain finite. The surface contribution has already
been taken into account. Therefore, the only contribution of the
electromagnetic field that survives in the quasihorizon limit is due
to the outer region. As we will see for the issue under discussion the
situation with quasi-black holes is very close to that with black
holes. We only repeat briefly the main standard steps that lead to the
corresponding expression. Consider the electromagnetic contribution to
the external mass, $M_{\mathrm{out}}^{\mathrm{em}}=
\int_{\mathrm{out}}(-T_{\quad 0}^{\mathrm{em}\,0}+T_{\quad
k}^{\mathrm{em}\,k})\,N\sqrt{g_{3}} \,d^{3}x$. Using the expression
for the electromagnetic field tensor $ T_{\quad \mu
}^{\mathrm{em}\,\nu }=\frac{1}{4\pi }(F_{\mu }^{\mu }F_{\nu }^{\mu
}-\frac{1}{4}\delta _{\mu }^{\nu }F_{\mu \nu }F^{\mu \nu })$, where $
F_{\mu \nu }=\frac{\partial A_{\nu }}{\partial x^{\mu
}}-\frac{\partial A_{\mu }}{\partial x^{\nu }}$ is the field tensor,
and $A_{\mu }$ is the four-potential, one can transform
$M_{\mathrm{out}}^{\mathrm{em}}$ into $M_{
\mathrm{out}}^{\mathrm{em}}=-\frac{1}{4\pi }\int_{\mathrm{out}
}F_{0k}F^{0k}\,N\sqrt{g_{3}}\,d^{3}x$. The next step consists in an
integration by parts, applying the Maxwell equation $\frac{\partial
_{\nu }(F^{\mu \nu }N\sqrt{g_{3}})}{N\sqrt{g_{3}}}=4\pi j^{\mu }$,
where $j^{\mu \text{ }}$ is the current, and the Gauss theorem. This
operation converts $ M_{\mathrm{out}}^{\mathrm{em}}$ into an integral
over a surface at infinity and a surface at the boundary of the
quasihorizon. The first contribution vanishes since, by assumption,
there are no currents at infinity. The second one gives us
$M_{\mathrm{out}}^{\mathrm{em}}=A_{0}=\varphi _{\mathrm{h}}Q \text{,
}$ where $Q$ is the charge enclosed within the quasihorizon, and $
\varphi _{h}$ is the electric potential on the horizon which is
uniform in the quasihorizon limit (see below). Thus, although the
expression $M_{ \mathrm{out}}^{\mathrm{em}}=\varphi _{\mathrm{h}}Q$
comes from the outer contribution, it simply reduces to a surface term
similarly to what happens in the black hole case \cite{bch,cart73}.

Now, we study in detail the behavior of the electric potential in the
quasihorizon limit, and show that on the quasihorizon $\varphi$
becomes constant. Indeed, in the derivation of the electromagnetic
contribution to the mass, we used the uniformity of the electric
potential in the quasihorizon limit, so that it can be pulled out of
the surface integral, so now we have to prove it. The proof can be
outlined in a way similar to the discussion of the surface gravity in
Sec. \ref{finiteness}. We require the finiteness of the
electromagnetic energy density which is equal to $\rho^{
\mathrm{em}\,}= -T_{\quad 0}^{\mathrm{em}\,0}=
\left(F_{0i}F_{0k}g^{ik}\right)/8\pi= \left(
F_{0l}^{2}+g^{ab}\frac{\partial \varphi }{ \partial
x^{a}}\frac{\partial \varphi }{\partial x^{b}} \right)/\left(8\pi
N^{2}\right)$. As the metric $g^{ab}$ is positive definite, all terms
enter this expression with the ``+'' sign, so that $
N^{-1}\frac{\partial \varphi }{ \partial x^{a}}$ should be
finite. Near the quasi-black hole limit, it is equal to
$N_{0}^{-1}\frac{\partial \varphi }{ \partial x^{a}}$ where
$N_{0}=N_{0}(x^{a})$ is the value of the lapse function on the
quasihorizon. Then, taking the limit $N_{0}\rightarrow 0$, we obtain
that the finiteness of $\rho $ entails that in the quasihorizon limit
$\frac{\partial \varphi}{\partial x^{a}}\rightarrow 0$, so that $
\varphi$ indeed becomes constant and can be indeed pulled out from the
surface integrals.

\subsection{The mass formula}

\subsubsection{The formula}

Putting all the masses together, the inner, the surface, and the outer
masses, we find that the total mass of a system containing a
quasi-blak hole is
\begin{equation}
M=\frac{\kappa A_{\mathrm{h}}}{4\pi}+
\varphi_{\mathrm{h}}Q+M_{\;\mathrm{out}
}^{\mathrm{matter}}\,.  \label{massformula}
\end{equation}
Note that for the extremal case, the term $\frac{\kappa
A_{\mathrm{h}}}{4\pi}$ goes to zero, since $\kappa$ is zero. Now,
Eq. (\ref{massformula}) is nothing else than the mass formula for
quasi-black holes and surroundings, which has precisely the same form
as the mass formula for black holes and surroundings
(\cite{bch,cart73,bard73,sm,fn}). In outer vacuum, where $M_{\;
\mathrm{out}}^{\mathrm{matter}}=0$, one has the mass of the
quasi-black hole is
\begin{equation}
M_{\mathrm{h}}=\frac{\kappa A_{\mathrm{h}}}{4\pi}+\varphi_{\mathrm{h}}Q\,.
\label{massformulastaticnothing}
\end{equation}
This is Smarr's formula \cite{sm} (see also \cite{fn}), but now for
quasi-black holes. Note that, if one considers a generic matter
configuration without a quasihorizon, the above arguments do not work
at all. So Eqs. (\ref{massformula}) and
(\ref{massformulastaticnothing}) are only valid for quasi-black holes,
and black holes.

\subsubsection{Example: spherically symmetric electrically charged
quasi-black hole}

Consider, as an example, the spherically symmetrical case when only
the electromagnetic field is present outside, so that the outer region
of the quasi-black hole is described by the Reissner-Nordstr\"{o}m
metric, i.e., $ds^2=-\left(1-2m/r+Q^2/r^2\right)dt^2+
dr^2/\left(1-2m/r+Q^2/r^2\right)+ r^2\left(d\theta^2 +\sin^2\theta
d\phi^2\right)$. Then, $A_{\mathrm{h}}=4\pi r_{\mathrm{h}}^2$,
$\kappa_{\mathrm{h}}=(r_{\mathrm{h}}-m)/ r_{\mathrm{h}}^2$,
$\varphi(r_{\mathrm{h}})=Q^2/r_{\mathrm{h}}$, and
$r_{\mathrm{h}}=m+\sqrt{ m^2-Q^2}$. Using
(\ref{massformula})-(\ref{massformulastaticnothing}) one finds
$M=M_{\mathrm{h}}=m$, as it should. This is valid in the non-extremal
as well as in the extremal cases. The extremal case is peculiar, since
for $m=Q=r_{\mathrm{h}}$, the surface contribution vanishes, and the
contribution for the mass is purely electromagnetic. It is instructive
to work out directly from Eqs (\ref{ma}) and (\ref{outerm}). Then,
$M_{\mathrm{surf}}=m-{ \ \ Q^2}/r_{\mathrm{h}}$, $M_{\mathrm{out}}=
{Q^2}/r_{\mathrm{h}}$, so that the total mass is equal to
$m$. Clearly, in the particular case of an extremal quasi-black hole,
again the surface contribution vanishes.

\subsubsection{Hairy properties of quasi-black holes: Mass, electric
potential, and charge}

Now, it is interesting to understand which quantities give hair and
which give no hair to the quasi-black holes. Let us start with the
mass. The inner mass properties discussed above show that there are
different quasi-black hole configurations characterized by the same
mass but different inner mass densities $T^0_0=\rho$ say. However,
this difference becomes negligible in the quasi-black hole limit since
$\rho$ is multiplied by the factor $N$ which, in this limit, vanishes
in the inner region. This means that the hairy remnants of the
original configuration, which exist in the mass density, are deleted
in the quasihorizon limit. It is also instructive to look at the
situation with the electromagnetic potential. From the definition of a
quasi-black hole, it follows that in the inner region the lapse
function $N$ goes as $N=\varepsilon f(x^{i})$, for some non-zero
well-behaved $f(x^{i})$, and with $\varepsilon \rightarrow 0$ in the
quasi-black hole limit. Consider a static distribution of charge. Then
the only non-zero component of the current $j^\mu$ is
$j^{0}=\frac{\rho _{ \mathrm{e}}}{N\sqrt{g_3}}$, where $\rho
_{\mathrm{e}}$ is the invariant electrical charge density. Then, it
follows that $F^{0k}\sim \varepsilon ^{-1}$ and so
$F_{k0}=\frac{\partial \varphi }{\partial x^{k}}\sim \varepsilon
$. Therefore, the potential in the inner region takes the form
\begin{equation}
\varphi =\mathrm{const}+\varepsilon h(x^{i})\,,
\label{potential}
\end{equation}
for some $h$. Thus,
$\varphi$ tends to a constant in the quasihorizon limit (see also the
discussion in \cite{extdust} on hairy properties for particular
spherically-symmetrical models). Again, as for the mass, the hairy
remnants of the original inner configuration, which exist in the
electric potential, are deleted in the quasihorizon limit. Finally,
let us see what happens to the the charge distribution. The situation
in this case somewhat different. The charge $Q$ is defined by
$Q=\int\rho _{\mathrm{e}} \sqrt{g_{3}}\,d^{3}x$, with no
multiplication by $N$ inside the integral. So different charge density
distributions can be considered as yielding some kind of
hair. However, to probe the corresponding details, an outer observer
should exchange information with an inner observer. But, this is
impossible because both the infinite tidal forces and the rescaling of
time coordinate used by the two observers, do not allow such an
exchange \cite{qbh}. Therefore, even if there is hair, it appears in
places that become unavailable for observations in the quasi-black
hole limit. In this respect, the transition from a black hole to a
quasi-black hole agrees with the no-hair theorems, thus extending
their meaning.

\subsubsection{Beyond the mass formula: corrections}

In the discussion above, we have already pointed that quasi-black and
black holes are distinct physical objects, and the method of
derivation of the formulas for both objects is different. This
difference is surely not revealed in the final mass formula
(\ref{massformula}) (or Eq. (\ref{massformulastaticnothing})). This,
in a sense, is quite natural since for an outer observer the
quasi-black hole is practically indistinguishable from a black hole
(stressing again that in general this is not so for the inner
region). However, the difference should be manifest in correction
terms which reflect how close the system is to the quasi-black hole
state.  Therefore, it is of interest to evaluate these corrections
which do not have an analogue for the black hole case. Besides, this
evaluation shows the accuracy of the formula. To this end, we discuss
the different contributions separately. For the inner and outer masses
the answer is simple. On the other hand, the surface contribution to
the mass and the term due to the charge and electric potential require
a careful evaluation.

According to the property (b) of quasi-black holes, $N\leq
N_{\mathrm{B} }\sim \varepsilon$ on the boundary near the
quasihorizon and in the inner region where one can write
$N=\varepsilon f(x^{i})$. Therefore, it follows from (\ref{inner})
directly that the correction $M_{\mathrm{in}}^{(1)}$ to the mass is
given by, $M_{\mathrm{in}}^{(1)}=O(\varepsilon )$. For the outer mass
there is no need in such an evaluation at all since there are no
specific features for a quasi-black hole there, it simply is given by
the contribution to the mass of the region outside the body's
boundary.

Let us now find the correction to the mass surface contribution, by
evaluating the main correction stemming from the term (\ref{al}),
responsible for the surface contribution. The \textquotedblleft-" term
is of the order $\varepsilon $ as it follows from the form of $N$ in
the inner region, listed above. To evaluate the \textquotedblleft+"
term, we consider first the non-extremal case. In the zeroth
approximation we can surely neglect the difference between a
quasi-black and a black hole in the \textquotedblleft+" term, and use
the expansion $N=\kappa l+O(l^{2})$, where $l$ is the proper distance
to the quasihorizon (or to the horizon in the black hole case). As a
result, $\left( \frac{\partial N}{\partial l} \right) _{+}=\kappa
+O(l)$, where $l\sim N\sim \varepsilon $. The evaluation for the
extremal and ultraextremal cases is similar. One only has to take into
account that, in the extremal case in (\ref{al}) one has $\left(
\frac{ \partial N}{\partial l}\right) _{+}\sim N\sim \exp
(-\frac{l}{l_{0}})\sim \varepsilon $. In the ultraextremal case the
corrections from the \textquotedblleft +" side turn out to be smaller
than $\varepsilon $, since in the limit $l\rightarrow \infty $, one
has $\left( \frac{\partial N}{ \partial l}\right) \sim \frac{N}{l}\sim
l^{-n-1}\sim \varepsilon ^{1+\frac{1}{n}}\ll N\sim \varepsilon $, for
some exponent $n$.

The correction connected with the surface contribution of the
electromagnetic field, $M_{\,\mathrm{surf}}^{(1)\mathrm{em}}$, is
given by 
\begin{equation}
M_{\,\mathrm{surf}}^{(1)\mathrm{em}}=\frac{1}{4\pi }\int d\sigma
_{0k}F^{0k}(\varphi -\varphi _{H})+\varphi _{H}\Delta q\,,
\label{emcor} 
\end{equation} 
where $d\sigma _{0k}=n_{k}\,d\sigma $ is
the standard surface element of a two-dimensional surface, $n_{k}$ is
the unit normal to the surface, $q=\frac{1}{4\pi }\int d\sigma
_{0k}F^{0k}$, $\Delta q$ is the charge enclosed between the
quasihorizon (or of the horizon in the black hole case) and the
boundary surface that approaches it, and $\varphi
_{H}=\mathrm{constant}$ is the value of the potential on the
quasihorizon (or on the horizon in the black hole case, but here the
difference between a black hole and quasi-black hole is
negligible). Using Eq. (\ref{potential}), which is valid in some
vicinity of the quasihorizon (or horizon in the black hole case) on
both sides, we see that the first term in (\ref{emcor}) is of the
order $ \varepsilon $. For the second term we can write, $\Delta q\sim
\Delta A$, where $\Delta A$ is the difference between the surface
areas. Then, $\Delta A\sim g^{ab}\Delta g_{ab}$. In the non-extremal
case $\Delta g_{ab}\sim l\sim N\sim \varepsilon $. In the extremal
case it follows from (\ref{ge}) that $\Delta g_{ab}\sim \exp
(-\frac{l}{l_{0}})\sim N\sim \varepsilon $.  In the ultraextremal
case, we have by definition $N\sim l^{-n}$ (see Sec. \ref{extrultra}) 
and
$\Delta g_{ab}\sim l^{-s}$, where $n>0$ and $s>0$ (a more detailed
discussion of the properties of such metrics is contained in
\cite{rom,vo,tr}).  Therefore, $\Delta g_{ab}\sim \varepsilon ^{p}$,
with $p= \frac{s}{n}$, and depending on the relation between $n$ and
$s$, both $p>1$ or $p<1$ are possible. Then,
$M_{\,\mathrm{surf}}^{(1)\mathrm{em}}=O(\varepsilon ^{p})$.

Thus, in brief, in the non-extremal and extremal cases all corrections
are of the order $\varepsilon$, whereas in the ultraextremal case the
several different corrections may contain $\varepsilon $ in different
powers as described above.

\subsubsection{Other black hole mimickers: Gravastars}

In this study of static spacetimes, we have mainly concentrated on
comparing aspects of quasi-black holes to true black holes. But there
are other interesting objects. Indeed, in recent years, there has been
some debate to what extent observational data can favor the existence
of black holes or can be ascribed to objects with size close to their
own gravitational radius but having no horizon, the black hole
mimickers \cite{mim}. Therefore, it is appropriate to compare
properties, such as the mass formula in the present article, of
quasi-black holes, not only to those of true black holes but also to
other types of mimickers. In this connection, we make some short
remarks about one of the most prominent mimickers, namely, gravastars
\cite{grav}. Their distinctive feature consists in that the almost
Schwarzschild-like outer metric is combined with a de Sitter-like
inner one.  Then, for our context, the difference between both types
of mimickers, i.e., quasi-black holes and gravastars, reveals itself
in the behavior of the lapse function in the inner region and on the
boundary surface. For gravastars, in contrast to quasi-black holes, in
the inner region $N$ does not vanish and is a monotonically decreasing
function of the radial coordinate. Therefore, it has a non-vanishing
derivative $\left( \frac{ \partial N}{\partial l}\right)_{-}$ on the
boundary, which, in turn, affects the mass value according to
Eq. (\ref{al}). As a result, both the inner region and surface
contribute to the mass, this contribution being model-dependent. For
quasi-black holes, as we have been discussing, all the information
about the inner region and boundary is deleted and the answer for the
mass formula has a universal form, just in the same spirit of black
hole physics. In this sense, quasi-black holes represent
configurations which are closer to black holes than gravastars, indeed
they are much better mimickers (although qualitative differences
between quasi-black holes and true black holes persist anyway, see
\cite{qbh} and \cite{mim}).

\section{Conclusions: The Abraham-Lorentz electron and other discussions}

\subsection{The Abraham-Lorentz electron and extremal quasi-black holes}

The above results have a rather unexpected implication concerning
another topic, namely, the problem of a self-consistent analogue of an
elementary particle in general relativity having a mass of pure
electromagnetic origin.  This is the correspondent to the
Abraham-Lorentz electron in flat spacetime.  In flat spacetime Coulomb
repulsion prevents such a construction, so one needs Poincar\'e
stresses for such a construction, But, by including gravitation, one
may possibly dispense with those stresses, the attractive force of
gravitation making the question reasonable within the theory of
general relativity. On a first glance, it would seem natural that, as
we want to have electromagnetic and gravitational forces alone, we
must require the absence of a bare tension on the surface. Otherwise,
this would mean that apart from electromagnetism and gravitation,
there were also external forces of different nature, of Poincar\'e
type, needed to keep the system in equilibrium. An attempt of this
kind was made on \cite{vf}, where it was argued that a charged shell
with empty space inside obeys this criteria in the extremal case,
$M=Q$. This was criticized in \cite{ep} where it was shown that,
actually, the surface stresses do not vanish in such a model even in
the extremal limit. Instead, another model was suggested in \cite{ep},
where the external extremal Reissner-Nordstr\"{o}m metric was glued to
the Bertotti-Robinson tube-like geometry inside. Then, it turned out
that the surface stresses vanish in the limit when the surface of
gluing approaches the horizon.

However, it follows from the results of the present article that the
two issues ``mass of pure electromagnetic origin'' and ``absence of
bare stresses'' in general relativity may be different in one
exceptional situation. If the surface of the charged body approaches
the quasihorizon, the contribution of the bare tension on the surface
to the total mass in the extremal case completely vanishes, although
these stresses by themselves remain finite. As a result, we obtain a
model in which a distant observer measures a mass as having purely
electromagnetic origin, although locally, on the surface there are
extraneous additional forces. Moreover, one can even allow
non-electromagnetic fields inside in the bulk, since anyway, their
contribution to the total mass vanishes in the quasihorizon limit. All
the region beyond the quasihorizon including the quasihorizon itself
turns out to be frozen and gives no contribution to the mass (for the
non-extremal case the inner region also is frozen but the boundary is
not).

It is also worth noting that the general statement of \cite{vf} about
the distinguished role of extremal black holes (now we would rephrase
it as \textquotedblleft black and quasi-black holes\textquotedblright)
turns out to be correct. They are suitable candidates for the role of
classical models of elementary particles since only in this case the
mass can have pure electromagnetic origin. Thus, in summary, as a
by-product, we have obtained that an extremal quasi-black hole (in
contrast to the non-extremal one) can serve as a physically reasonable
classical model of an Abraham-Lorentz electron in that both the inner
and surface contribution of forces with non-electromagnetic origin
vanish. In doing so, we showed that one may weaken the requirement of
vanishing surface stresses since the finite stresses have zero
contribution to the total mass.

\subsection{Other discussions}

We have traced how the limiting transition from the static
configuration to the quasi-black hole state reveals itself in the mass
formula. It turned out that there is a perfect one-to-one
correspondence between the different contributions for the total mass
of a quasi-black hole and the mass formula for black holes. In
particular, the inner contribution to the total mass vanishes in the
quasi-black hole limit, and surely it is absent in the black hole case
from the very beginning. The contribution of the surface stresses in
the quasi-black hole corresponds just to the contribution from the
horizon surface of a black hole. This is non trivial, since the
corresponding terms have quite different origins. In the quasi-black
hole case they are due to the boundary between both sides of the
surface. In the black hole case only one side, the external, is
relevant and the integrand over this surface has nothing to do with
the expression for surface stresses. Nonetheless, both terms coincide
in the limit under discussion.

The essential difference between non-extremal and extremal quasi-black
holes consists in that the first case the surface stresses become
infinite but have finite contribution to the total mass, while in the
second case they are finite but have zero contribution to the total
mass. Actually, we extended in the present paper the notion of a
quasi-black hole admitting infinite surface stresses. As far as the
mass is concerned, in the non-extremal case the surface of a
quasi-black hole reveals itself in a way similar to a membrane in the
membrane paradigm setup \cite{membr}, whereas in the extremal one we
have \textquotedblleft membrane without membrane\textquotedblright,
paraphrasing famous Wheeler's remarks \cite{wheeler}.  By itself, the
system with infinite stresses looks unphysical and this was the reason
why non-extremal quasi-black holes were rejected in
\cite{qbh}. Nonetheless, consideration of such systems has at least
methodical interest since it helps to understand better the
relationship between quasi-black holes and black holes and the
distinction between non-extremal and extremal limits in this
context. In particular, it is of interest to trace the similarity and
distinction between quasi-black holes and black holes from the
viewpoint of the membrane paradigm in a more general setting.

Of course, by adding rotation all these matters may become even more
interesting.

\begin{acknowledgments}
O. Z. thanks Centro Multidisciplinar de Astrof\'{\i}sica -- CENTRA for
hospitality and a stimulating working atmosphere. This work was
partially funded by Funda\c c\~ao para a Ci\^encia e 
Tecnologia (FCT) - Portugal, through project POCI/FP/63943/2005.
\end{acknowledgments}


\end{document}